\documentclass[journal=nalefd,manuscript=letter]{achemso}

\usepackage[version=3]{mhchem} 

\setkeys{acs}{maxauthors=99, etalmode=truncate}

\author{Gerard N. Hinsley}
\email{gerard.hinsley@desy.de}
\author{Fabian Westermeister}
\author{Bihan Wang}
\author{Kuan Hoon Ngoi}
\author{Shweta Singh}
\author{Rustam Rysov}  
\author{Michael Sprung}
\affiliation[DESY]
{Deutsches Elektronen-Synchrotron DESY, Notkestr. 85, 22607 Hamburg, Germany}
\author{Cameron M. Kewish}
\affiliation[ANSTO]
{Australian Nuclear Science and Technology Organisation, Australian Synchrotron, Victoria 3168, Australia}
\alsoaffiliation[La Trobe]
{Department of Mathematical and Physical Sciences, La Trobe University, Bundoora, Victoria 3086, Australia}
\author{Grant A. van Riessen}
\affiliation[La Trobe University]
{Department of Mathematical and Physical Sciences, La Trobe University, Bundoora, Victoria 3086, Australia}
\alsoaffiliation[MCN]
{Melbourne Centre for Nanofabrication, Clayton, Victoria 3168, Australia}

\author{Ivan A. Vartanyants}
\email{ivan.vartaniants@desy.de}
\affiliation[DESY]
{Deutsches Elektronen-Synchrotron DESY, Notkestr. 85, 22607 Hamburg, Germany}

\title[An \textsf{achemso} demo]
   {Dynamic X-ray coherent diffraction analysis: bridging the timescales between imaging and photon correlation spectroscopy}

\abbreviations{CXDI, XPCS,SNR}

\begin{document}

\begin{abstract}
  The advent of diffraction limited sources and developments in detector technology opens up new possibilities for the study of materials \textit{in situ} and \textit{operando}. Coherent X-ray diffraction techniques such as coherent X-ray diffractive imaging (CXDI) and X-ray photon correlation spectroscopy (XPCS) are capable for this purpose and provide complimentary information, although due to signal-to-noise requirements, their simultaneous demonstration has been limited.
  Here, we demonstrate a strategy for the simultaneous use of CXDI and XPCS to study \textit{in situ} the Brownian motion of colloidal gold nanoparticles of 200~nm diameter suspended in a glycerol-water mixture. 
  We visualise the process of agglomeration, examine the spatiotemporal space accessible with the combination of techniques, and demonstrate CXDI with 22~ms temporal resolution.
\end{abstract}


~\\
Understanding the behaviour and function of materials at the nanoscale necessitates the use of \textit{in situ} and \textit{operando} characterisation approaches. 
Techniques using coherent X-ray illumination are becoming increasingly capable for the purpose of characterizing nanostructures \textit{in situ} with diffraction limited sources~\cite{weckert_potential_2015,schroer_petra_2018} and the development of new fast-framing detectors~\cite{bronnimann_hybrid_2018,donath_eiger2_2023}.

Two complimentary techniques, X-ray photon correlation spectroscopy (XPCS)~\cite{shpyrko_x-ray_2014,lehmkuhler_femtoseconds_2021} and coherent X-ray diffraction imaging (CXDI)~\cite{miao_beyond_2015}, can take advantage of these improvements for the analysis of sequentially measured coherent diffraction patterns.
XPCS analyses the correlation in the measured intensities $I(q,t)$ at time $t$ and after some delay $\tau$ for a given scattering vector $q$, and has been widely used to study \textit{in situ} and \textit{operando} dynamics~\cite{steinruck_concentration_2020,otto_x-ray_2022,
leheny_rheo-xpcs_2015,johnson_operando_2019,yavitt_revealing_2020}. 
The information obtained by XPCS, however, is the average over the entire illuminated region, and as such is unable to provide real-space structural information.

CXDI, alternatively, uses iterative phase-retrieval algorithms to  provide real-space images of the object~\cite{miao_beyond_2015}.
The inversion process of these diffraction patterns, however, is quite difficult, typically requiring a significantly higher signal-to-noise ratio (SNR) than XPCS for the algorithms to convergence to an accurate solution. This limitation has hindered the possibility to combine the two techniques simultaneously, and has driven interest in the development of novel strategies to improve the reconstruction.
One of these methods, Ptychography, introduces translational diversity in a set of diffraction patterns, where the overlap between adjacent frames can be exploited as a strong constraint allowing many experimental constraints to be relaxed. 
Consequently, ptychography has been attractive to use for the study of many systems \textit{in situ}~\cite{grote_imaging_2022,baier_situ_2016,baier_stability_2017, patil_x-ray_2016,hoydalsvik_situ_2014}, however this requirement of a high degree of spatial overlap introduces a trade-off between temporal resolution and the scanning area.
As such, alternative strategies have been developed to improve the robustness of CXDI for imaging dynamic systems~\cite{hinsley_dynamic_2020,hinsley_towards_2022,lo_situ_2018,takayama_dynamic_2021,tao_spatially_2018}.
This has resulted in the first demonstration of the complimentary information obtained by XPCS and CXDI which reached a temporal resolution of 0.1~s~\cite{takazawa_coupling_2023} - an order of magnitude faster than the 1.2~s achieved using ptychography~\cite{deng_velociprobe_2019}.
The experimental arrangement used in this high temporal resolution CXDI demonstration, however, required the use of a triangular aperture and the \textit{a priori} characterisation of the probe function through a ptychography measurement.

Here, we demonstrate how these results could be improved through a combination of XPCS and CXDI data analysis methods applied to an extended time-series acquisition of high frame-rate data.
By first reconstructing data integrated over many diffraction patterns, we obtain images of a blurred object with corresponding support functions. 
The CXDI reconstructions are then refined by using these supports as an initial guess for the reconstruction of smaller subsets of data, resulting in clearer images with high temporal resolution. 
This approach then allows for the simultaneous application of CXDI and XPCS while requiring no reference aperture or knowledge of the probe beam.
We use this approach to study the Brownian motion of Au colloidal nanoparticles with a diameter 200~nm and 300 kDa Poly(ethylene glycol) methyl (PEG) ligand, suspended in a dilute glycerol-water mixture inside a thin capillary.
We demonstrate how performing these measurements simultaneously provides complimentary information on the nature of the dynamics, while also demonstrating CXDI with a temporal resolution of 22~ms.

Data were collected at the P10 Coherence Applications beamline at PETRA III, where 21,000 coherent diffraction patterns were recorded using an EIGER 4M detector at a rate of 0.714~kHz (total acquisition time $\approx 30$~s). 
We studied the dynamics at two different temperatures, 300~K and 340~K, in order to examine the accessible spatiotemporal space by CXDI.
In-depth experimental details are provided in the Supplementary Information (SI).


\begin{figure}
    \centering
    \includegraphics[width=0.5\textwidth]{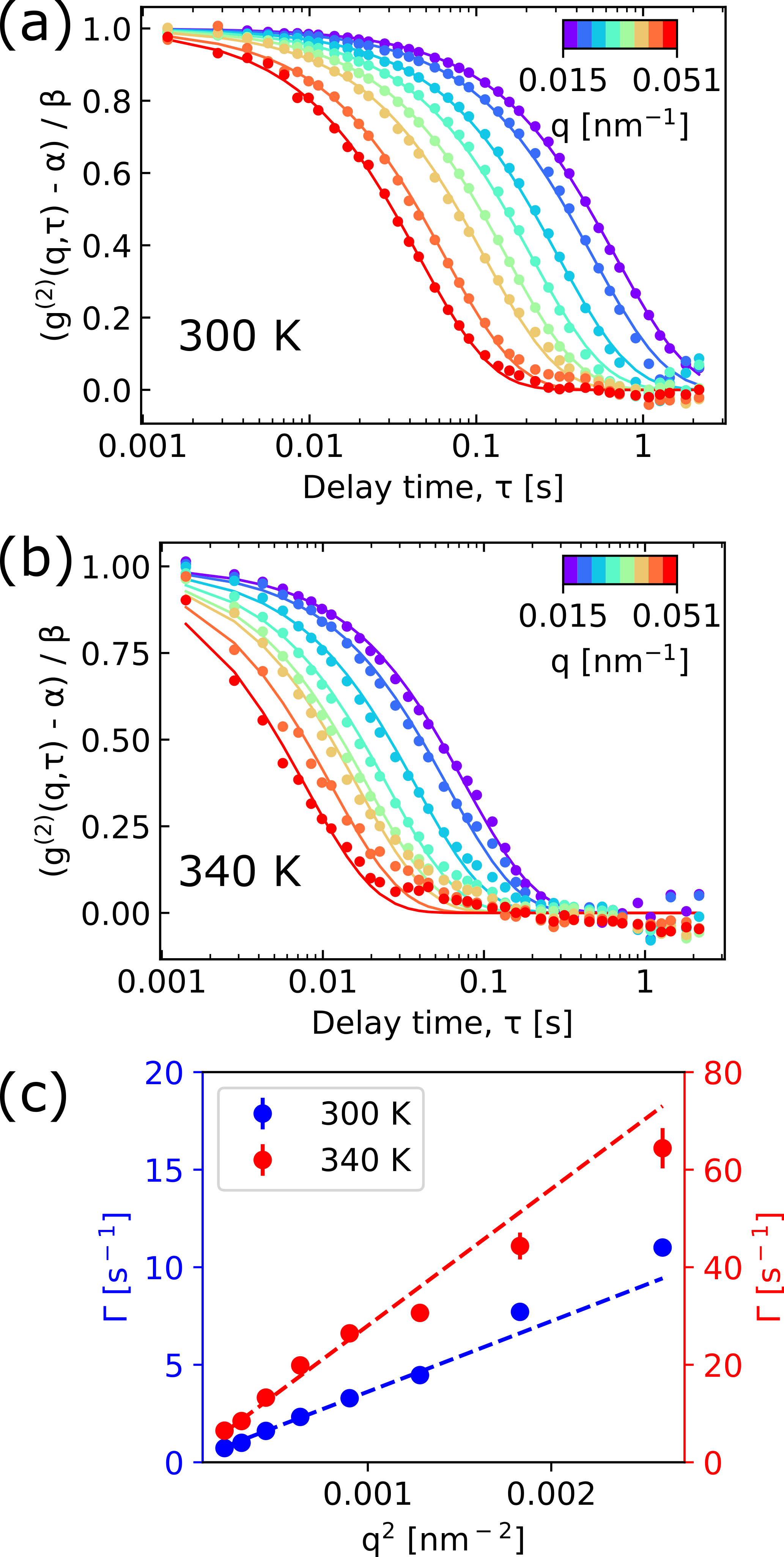}
    \caption{
    Calculated $g^{(2)}(q,\tau)$ functions (points) and their fits (solid lines) as a function of delay time at temperatures of 300~K (a) and 340~K (b) over $3$~s of data near the beginning of the series. 
    The colour represents the eight different $q$-partitions used for the analysis, where the partitions are spaced with equal d$q/q$ steps. 
    (c) Plot of $\Gamma$ as a function of $q^2$ for both temperatures. Linear fits are indicated by the dashed lines, and provide values of: $D_{\textnormal{300}}=3,618 \pm 18$~nm$^2$/s and $D_{\textnormal{340}}=28,024 \pm 423$~nm$^2$/s.
    }
    \label{fgr:XPCS-fit}
\end{figure}

To analyse the data using XPCS, the normalised intensity autocorrelation function, $g^{(2)}(q,\tau)$, is calculated by

\begin{equation} \label{eqn:XPCS g2 long}
    g^{(2)}(q,\tau) = \frac{\left \langle I(q,t) I(q,t+\tau) \right \rangle}
                         {\left \langle I(q,t)  \right \rangle ^2}
                         = 1 + \beta \left| g^{(1)}(q,\tau) \right|^2,
\end{equation}

\noindent where $\beta$ is the speckle contrast, and $g^{(1)}(q,\tau)$ is the intermediate scattering function.
For a system exhibiting Brownian motion, $g^{(1)}(q,\tau)$ can be described by~\cite{lehmkuhler_femtoseconds_2021}

\begin{equation} \label{eqn: KWW 1}
    |g^{(1)}(q,\tau)|^{2} = \exp{[-2(\Gamma \tau)^{\gamma}]},
\end{equation}

\noindent where $\Gamma=Dq^2$ is the relaxation rate with $D$ being the diffusion coefficient of the particles, and $\gamma$ is a measure of the distribution of relaxation times. 
We fit this data using

\begin{equation} \label{eqn: XPCS fit}
    g^{(2)}(q,\tau) = \alpha + \beta \left( \exp{[-2(\Gamma \tau)^{\gamma}] }  \right),
\end{equation}

\noindent where $\alpha$ represents the baseline, and we set $\gamma = 1$ as a fixed parameter as the motion of the nanoparticles is expected to be Brownian. Fits with $\gamma$ as a free parameter are shown in the SI.

Figure~\ref{fgr:XPCS-fit}(a,b) presents the results of this analysis performed over 3~s near the beginning of each dataset at temperatures of 300~K and 340~K, respectively. The solid points represent values of $g^{(2)}(q,\tau)$ calculated by Eq.~\eqref{eqn:XPCS g2 long}, and were normalised by subtracting the baseline and division by the contrast value obtained from the fits. The solid lines in Figs.~\ref{fgr:XPCS-fit}(a,b) are the fits for each $q$ partition obtained by Eq.~\eqref{eqn: XPCS fit}.
These results clearly indicate the expected behaviour where the effect of increasing the temperature of the system translates to a decay in the correlation at shorter delay times.
The XPCS results for the whole time series, as well as further details of the fitting procedure, are included in the SI.

Figure~\ref{fgr:XPCS-fit}(c) shows a plot of the values of $\Gamma$ obtained from the fitting procedure obtained at both temperatures. 
By calculating $d\Gamma / dq^{2}$, the slope of the results in Fig.~\ref{fgr:XPCS-fit}(c), we obtain estimates of the diffusion coefficient of $D^{\textnormal{XPCS}}_{\textnormal{300}}=3,618 \pm 18$~nm$^2$/s and $D^{\textnormal{XPCS}}_{\textnormal{340}}=28,024 \pm 423$~nm$^2$/s.
With the results obtained at 300~K, we perform a micro-rheology analysis (details in the SI) and estimate that the solution contains approximately 88\% glycerol and 12\% water. 
It follows that for a solution with this composition at a temperature of 340~K, we expect that $D=55,102$~nm$^2$/s. This value is about twice the actual value obtained, indicating that our nanoparticles are less mobile than expected.
This sub-diffusive behaviour indicates that the system is either at a lower temperature than 340~K, or possibly there is also some degree of agglomeration of the nanoparticles. 
Understanding the origin of the sub-diffusive behaviour can be revealed by analysing the spatial distribution of particles by CXDI.


\begin{figure}
    \centering
    \includegraphics[width=\linewidth]{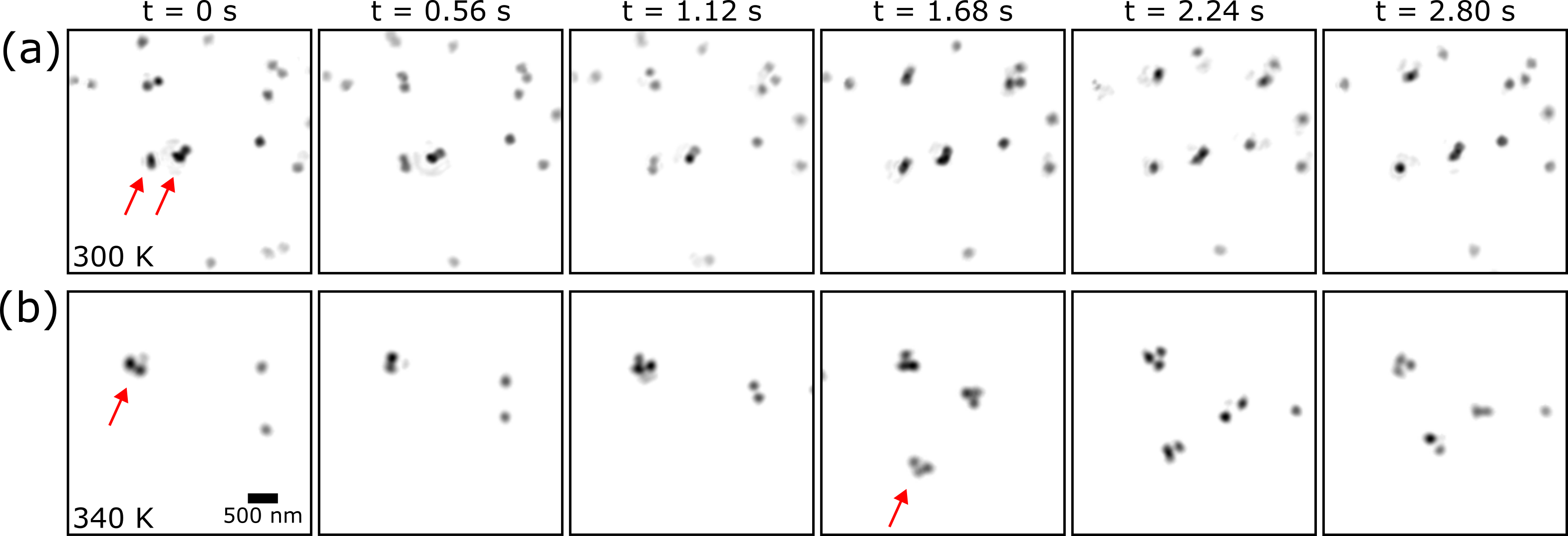}
    \caption{Representative reconstructed amplitude images at 300~K (a) and 340~K (b), over the same temporal range as the XPCS analysis shown in Fig.~\ref{fgr:XPCS-fit}. Images were cropped to a size of $4 \times 4$~$\mu$m$^{2}$ around the centre. 
    Before reconstruction, data were summed over 50 frames (300~K) and 16 frames (340~K) which correspond to temporal resolutions of 70~ms and 22~ms, respectively. The red arrows point to the presence of agglomerations. The scalebar applies to all images.
    }
    \label{fgr:CXDI}
\end{figure}

As the SNR of a single diffraction pattern is too low to reconstruct by CXDI, data were summed over multiple diffraction patterns using a sliding temporal window (see SI) in order to enable reconstruction convergence. 
Reconstructable datasets for both temperatures were initially generated by summing together 50 diffraction patterns, corresponding to a temporal resolution of 70~ms.
These datasets were then reconstructed using PyNX~\cite{favre-nicolin_pynx_2020}, where further details can be found in the SI.
For the data at 340~K, a second dataset was then generated by summing together 16 diffraction patterns, corresponding to a temporal resolution of 22~ms.
In the reconstruction of the second dataset, the final support functions from the first dataset were used as an initial guess.
Representative reconstructed amplitude images, cropped to a size of $4 \times 4$~$\mu$m$^{2}$ around the centre and extracted over the same temporal range as the results shown in Fig.~\ref{fgr:XPCS-fit}, are shown in Fig.~\ref{fgr:CXDI}(a,b) for the temperatures of 300~K and 340~K, respectively. Further details on the data pre-processing and reconstruction steps, as well as reconstructed movies of the datasets, can be found in the SI.

For both temperatures, we can identify the presence of individual nanoparticles as well as agglomerates (indicated by the red arrows in Fig.~\ref{fgr:CXDI}), all of which undergo Brownian motion. For the 300~K case, there exists two agglomerations consisting of two particles, with the rest of the particles being individual and no particle-particle interactions occurring.
In contrast, at 340~K, we not only see faster movement from the particles and the presence of multiple agglomerations, but additionally we can see the particles aggregating over time (see SI movie 2). 
Near the beginning of the measurement there exists two three-particle agglomerates, but after $\approx 20$~s there appears to be two large agglomerations consisting of roughly seven to eight nanoparticles each. 
The existence of these agglomerations directly leads to a reduction in the diffusion coefficient, which contributes to the explanation of the discrepancy between the calculated and observed diffusion coefficients obtained by XPCS at 340~K.

For PEGylated Au nanoparticles in glycerol-water mixtures, it has been observed that glycerol can outcompete PEG for the available water molecules in the system, leading to attractive PEG-PEG interactions, which then results in nanoparticle agglomeration~\cite{jain_three-step_2022,schulz_situ_2024}.
This attraction between the PEG ligands can explain the existence of agglomerations at both temperatures observed here.
Further study is required to understand in more detail the agglomeration process observed at 340~K, in particular, whether the elevated temperature or radiation damage effects cause the process to be enhanced.

The correlations obtained by the XPCS analysis for the full 340~K dataset (see SI) shows no clear indication of the onset of the agglomeration process. This highlights the complimentary nature of combining both CXDI and XPCS simultaneously. These results have implications for the use of nanoparticles as contrast agents in biological environments~\cite{pelaz_diverse_2017, feliu_vivo_2016}, and particularly for examining induced agglomeration processes~\cite{otto_x-ray_2022}.

\begin{figure}
    \centering
    \includegraphics[width=\linewidth]{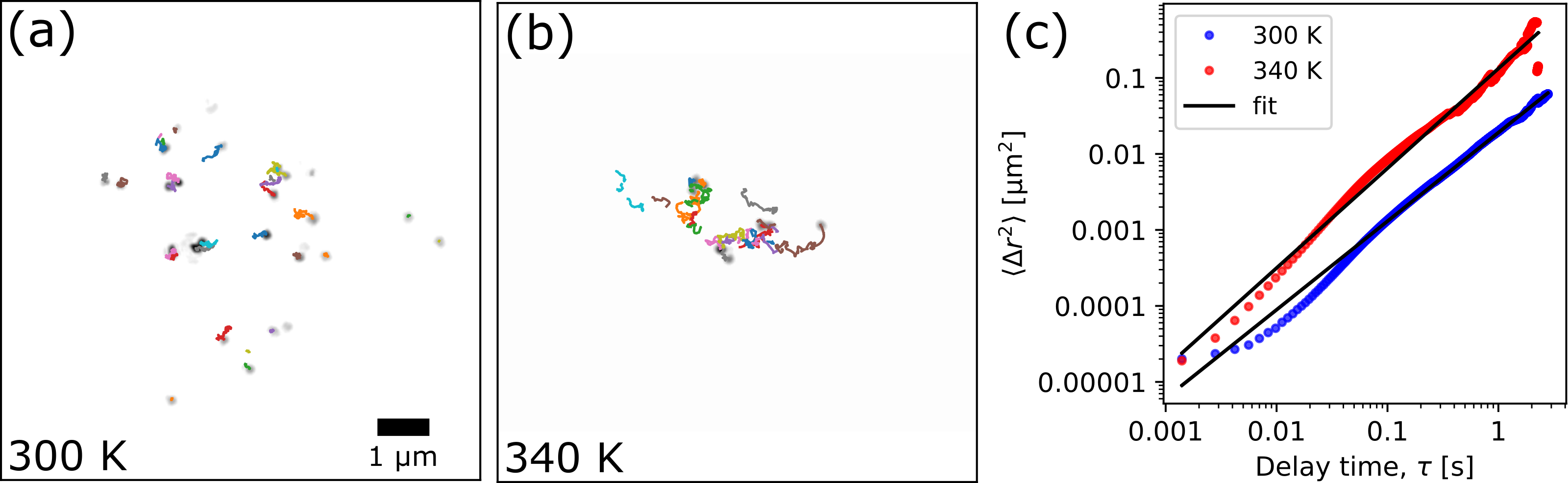}
    \caption{Single particle tracking results show the motion of the individual particles over 3~s at the beginning of the dataset at temperatures of 300~K (a) and 340~K (b). (c) The gradient of the ensemble mean-squared displacement provides estimates of the diffusion coefficient of $D_{\textnormal{300}} =  4,773 \pm 1,223$~nm$^2$/s, and  $D_{\textnormal{340}} =  33,222 \pm 15,053$~nm$^2$/s. The uncertainty in the SPT results was estimated using the standard error.
    The scalebar applies to both images.
    }
    \label{fgr:SPT}
\end{figure}

The dynamics identified in the CXDI reconstructions can be better understood by performing single particle tracking (SPT) analysis, where particle tracking algorithms can identify the particle locations for each point in the time-series~\cite{crocker_methods_1996}. 
The mean-squared displacement (MSD), $\langle \Delta r^2 \rangle$, of the individual particles can be related to the diffusion coefficient by~\cite{crocker_methods_1996} $\langle \Delta r^2 \rangle = 4 D \tau$, providing a comparison to those obtained by XPCS. 
The degree of uncertainty was estimated using the standard error~\cite{catipovic_improving_2013} which is dependent on the number of tracked particles.

Figure~\ref{fgr:SPT} shows the SPT results extracted over the same time interval as shown in Fig.~\ref{fgr:CXDI}, where details on the parameters and error estimation are found in the SI.
The obtained estimates of the diffusion coefficients are $D^{\textnormal{CXDI}}_{\textnormal{300}}=4,773 \pm 1,223$~nm$^2$/s, and $D^{\textnormal{CXDI}}_{\textnormal{340}}=33,222 \pm 15,053$~nm$^2$/s. 
These values are within error compared to those obtained by XPCS, but are slightly overestimated. Overestimation of the values can be partially attributed to a number of factors such as misalignment between frames, missing particles in the reconstruction, or blurred motion leading to misidentified particle positions.

\begin{figure}
    \centering
    \includegraphics[width=\linewidth]{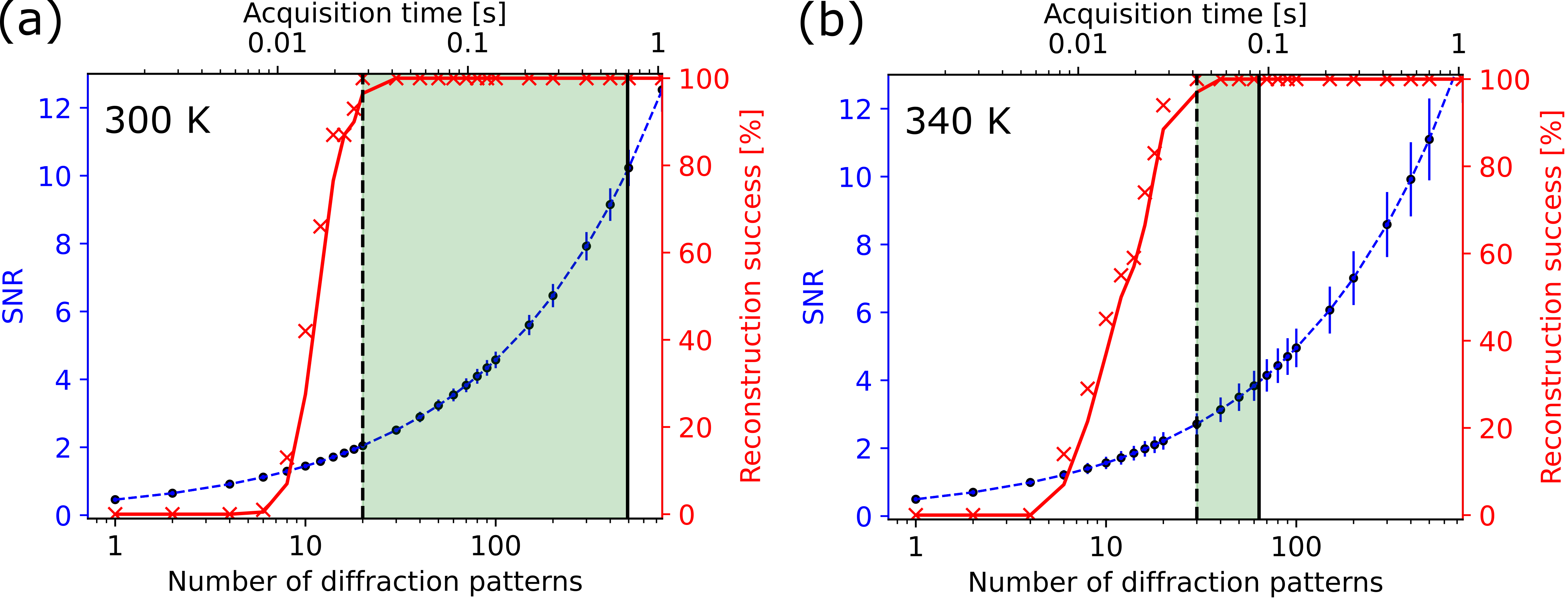}
    \caption{Signal-to-noise ratio analysis for reconstructing time-series CXDI datasets at 300~K (a), and 340~K (b).
    The black dashed line represent the limits to the accessible time range, where at 300 K, $f_{\textnormal{SNR}}=2.0$, and at 340~K, $f_{\textnormal{SNR}}=2.7$. The critical sampling frequency, $f_c$, is marked by a solid black line. The green shaded region between these points represents the temporal space accessible by CXDI.
    }
    \label{fgr:SNR}
\end{figure}

The main consideration in the accuracy of the SPT analysis revolves around the relationship between the quality of the reconstructions, the temporal resolution and SNR, as well as the degree of dynamics of the sample.
To investigate the relationship further between these variables and more generally the limits on the ability to perform dynamic CXDI experiments, we analysed the reconstruction quality as a function of SNR.
This was achieved by first generating multiple datasets each of which consist of the integration over a varying number of diffraction patterns using a sliding temporal window, calculating the SNR using Eq.~(S5) in the supplementary information, and then performing reconstructions of each dataset. 
The results are shown in Fig.~\ref{fgr:SNR}, where further details of the SNR calculations and reconstruction process are found in the SI. 

The blue curve in Fig.~\ref{fgr:SNR} represents the SNR value which increases with an increasing number of summed frames while the red curve represents the percentage chance that a reconstruction will converge. 
The solid vertical black lines represent the critical sampling frequency, $f_c$, which are the points at which the integration of diffraction patterns is too high such that the object dynamics may become blurred. These were estimated to be $f_c = 1.45$~Hz (300~K) and $f_c = 11.21$~Hz (340~K). These values represent an upper limit to the integration and are dependent on the sample dynamics (see SI for how this value was estimated).
At low numbers of summed diffraction patterns, there is not enough signal to allow the iterative algorithms to consistently converge to a reasonable solution. As the number of summed diffraction patterns increases, there becomes a SNR value which results in a converged reconstruction 100\% of the time. This point is taken as a lower limit for reconstructing dynamic CXDI data, and is indicated in Fig.~\ref{fgr:SNR} as a vertical black dashed line. 
We see that this limit is reached at a SNR around 2.0 to 2.7, corresponding here to temporal resolutions of 28~ms and 42~ms, respectively.
While it is possible to obtain a reconstruction integrating a smaller number of diffraction patterns, an increasingly large amount of time is required to obtain a good reconstruction. 
For very large datasets, which will soon become commonplace, this becomes unreasonable using standard approaches.
Future efforts into developing more robust CXDI reconstruction approaches with low SNR is required to push the ability to reconstruct data below this limit.


To summarise, we have demonstrated simultaneous CXDI and XPCS to study the Brownian motion of colloidal gold nanoparticles with a temporal resolution of 22~ms. 
The simultaneous analysis allows complimentary information to be extracted, where we observed Brownian motion at temperatures of 300~K and 340~K, as well as the process of agglomeration at a temperature of 340~K.

As the scattering power of the sample directly relates to the ability to reconstruct a diffraction pattern, we investigated the effect of SNR on the reconstruction performance to provide an estimate of the spatiotemporal space accessible for CXDI. We determine that a lower limit is reached at SNR~$\approx~2$. This can act as a guide for the design of future \textit{in situ} and \textit{operando} experiments to study a wide range of dynamic systems.

The results demonstrated here were enabled by recent advancements in hybrid photon counting detectors technology, namely; fast frame rate, double-buffering to readout frames with zero deadtime, high dynamic range from single photon counting to over 1e6 photons/pixel/sec. 
To image dynamic systems using coherent diffractive techniques on faster timescales will require future improvements in detector technology, to take advantage of the increased brightness at diffraction limited light sources, or further development of novel reconstruction approaches (e.g., Ref.~\cite{hinsley_towards_2022}).

\begin{acknowledgement}
The authors thank Stephan Roth, JunGui Zhou, and Yingjian Guo.

\end{acknowledgement}

\begin{suppinfo}
Reconstructed amplitude movies for both 300~K and 340~K data (.mp4).
Additional experimental details, and detailed data processing steps for XPCS, CXDI, and SPT analyses.
\end{suppinfo}

\renewcommand{\thefigure}{S\arabic{figure}}
\renewcommand{\theequation}{S\arabic{equation}}
\setcounter{figure}{0}
\setcounter{equation}{0}

\section{Supplementary information}

\section{Sample description}

Au nanoparticles were obtained from Nanopartz (Loveland, US)~\cite{noauthor_gold_nodate}. The nanoparticles have a diameter of 200~nm, are functionalised with 300~kDa Poly(ethylene glycol) methyl, and were dispersed in a solution of 18MEG DI water.
This acts as the stock solution, where the weight concentration of nanoparticles is 3.9 mg/mL, and the volume fraction is approximately 0.02\%.

A volume of 100~$\mu$L of the stock solution was extracted and centrifuged at 5000 rpm for 5 minutes. The liquid was then aspirated, and the same volume of glycerol was added. This was then sonicated to redisperse the particles. 
From this, 10~$\mu$L of the solution was extracted and combined with 90~$\mu$L of glycerol, diluting the solution, before the sample was re-sonicated to again disperse the nanoparticles uniformly within the solution.

Rectangular capillaries with dimensions 0.5~mm~$\times$~5~mm~$\times$~0.05~mm (H~$\times$~V~$\times$~W) were then filled with the solution, and sealed to enable vacuum compatibility.

\section{Experiment and data pre-processing}

\begin{figure}
    \centering
    \includegraphics[width=\linewidth]{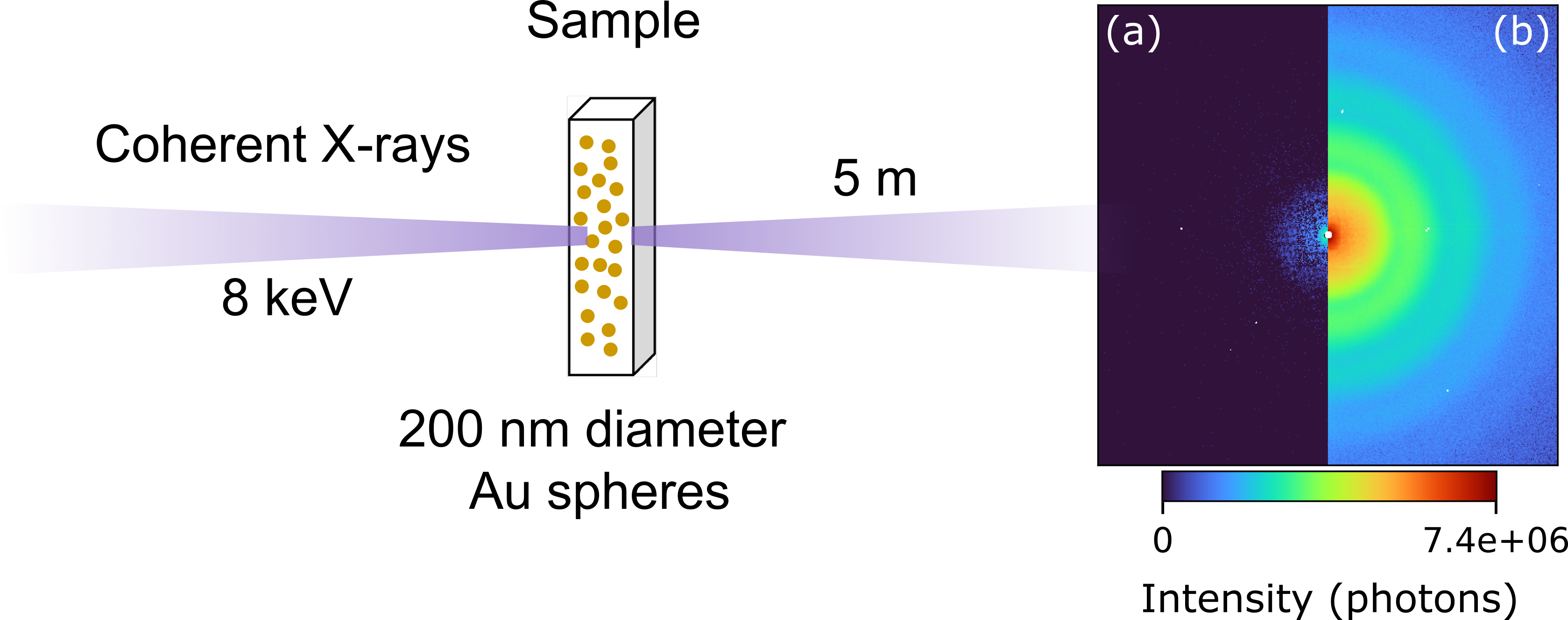}
    \caption{A diagram of the experimental setup, where coherent X-rays with an energy of 8~keV are focussed to a size of 2.8~$\times$~2.3~$\mu$m$^2$ (H~$\times$~V) at the focal plane. The capillary containing the dispersion of the 200~nm diameter Au nanoparticles was positioned at the focal plane. Diffraction patterns were measured 5~m downstream of the sample position. The diffraction pattern shows one acquisition (a) and the summation of all diffraction patterns (b).
    }
    \label{fgr:Experiment}
\end{figure}

The experiment was performed at the P10 Coherence Applications beamline at PETRA III, as shown in Fig~\ref{fgr:Experiment}. 
The coherent source size of the incident 8~keV X-rays was controlled by apertures which were set to a size of 80~$\times$~125~$\mu$m$^2$ (H~$\times$~V). This was then focussed using compound refractive lenses to a size of 2.8~$\times$~2.3~$\mu$m$^2$ (H~$\times$~V) at full width half maximum (FWHM). The capillary containing the nanoparticles was placed at the focal plane, and the detector was placed 5~m downstream of the sample.
Diffraction patterns were recorded at 0.714~kHz using an EIGER 4M detector, which has pixels of size 75~$\times$~75~$\mu$m$^2$. 
A tungsten cylinder was used as a beamstop to block the central beam, and was attached to a semi-transparent beamstop consisting of two Si wafers of dimensions 3~mm~$\times$~3~mm~$\times$~100~$\mu$m, and 5~mm~$\times$~5~mm~$\times$~100~$\mu$m (H~$\times$~V~$\times$~W).

\begin{figure}
    \centering
    \includegraphics[width=0.75\linewidth]{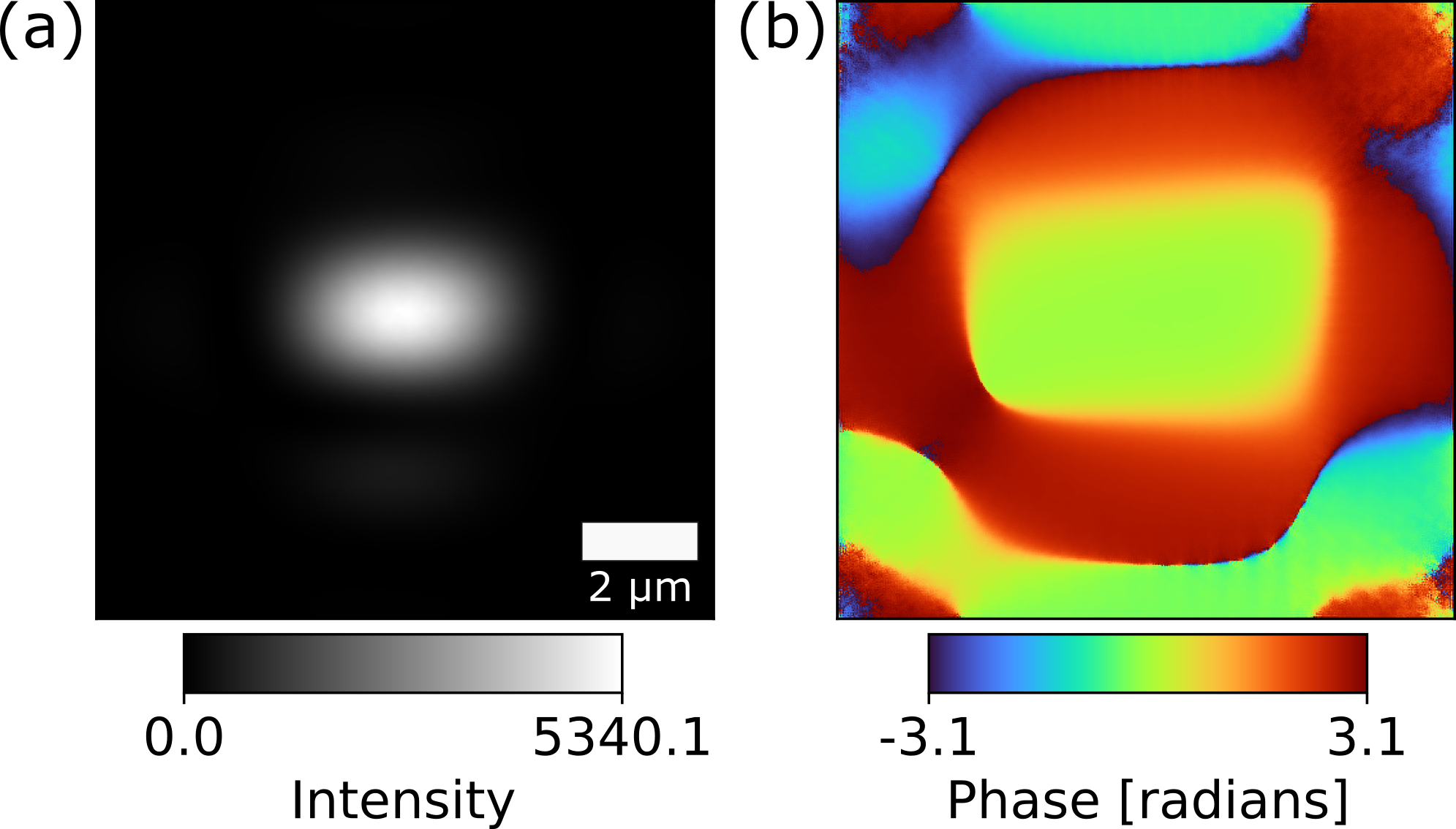}
    \caption{ 
    The reconstructed probe intensity (a) and phase (b) obtained from a ptychography measurement using the same experimental geometry. For this data, the FWHM of the probe intensity is 1.7~$\mu$m~$\times$~2.8~$\mu$m (H~$\times$~V).
    }
    \label{fgr:Reconstructed probe}
\end{figure}

A representative image showing the expected shape of the probe intensity and phase at the focal plane is displayed in Fig.~\ref{fgr:Reconstructed probe}. This probe function was retrieved from a ptychography measurement during a previous experiment using the same experimental setup, where no aperture or special experimental geometry is required.

\begin{figure}
    \centering
    \includegraphics[width=\linewidth]{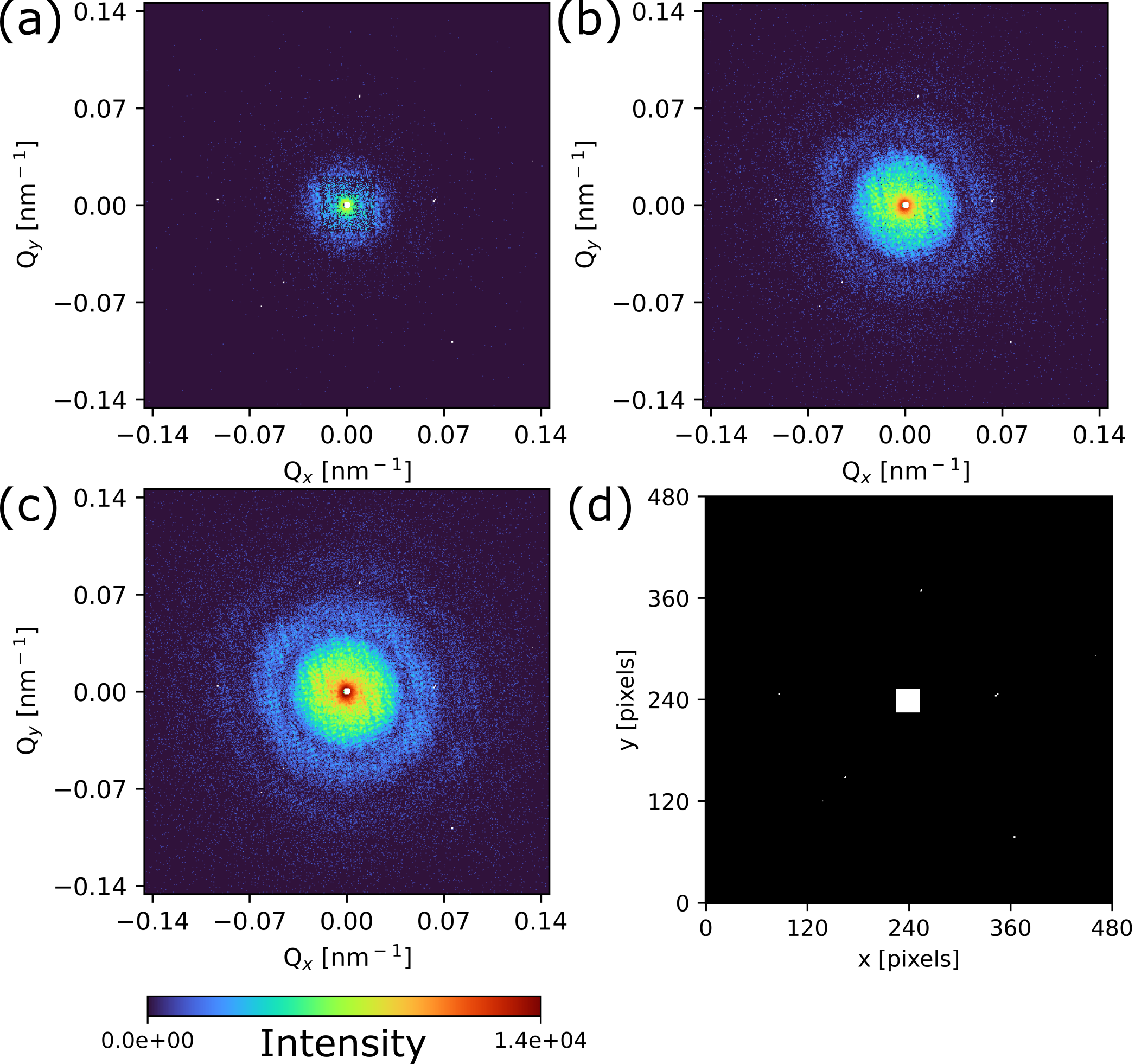}
    \caption{
    (a)-(c) Images of the recorded diffraction patterns after post-processing. (a) a single diffraction pattern from the 300~K data. (b) Diffraction pattern after summing together 16 frames. (c) Diffraction pattern after summing together 50 frames.
    (d) The pixel mask used in the reconstructions. White pixels in (a)-(c) represent bad pixels with incorrect intensities, while (d) includes extra pixels which were added to additionally mask intensities related to the probe. Masked pixels were allowed to fluctuate freely in the reconstruction.
    }
    \label{fgr:SNR diff}
\end{figure}

Data were prepared for reconstruction by cropping diffraction patterns to a size of 480~pixels~$\times$~480~pixels around the beam centre. The intensities of the pixels affected by the semi-transparent beamstop were then scaled by multiplying the values by the expected X-ray absorption at 8 keV. 
A representative image of a single diffraction pattern after pre-processing is shown in Fig.~\ref{fgr:SNR diff}(a), while (b) and (c) show diffraction patterns after summing together 16 and 50 frames, respectively.
A pixel mask was simultaneously created to remove dead pixels, and the effect of the tungsten beamstop for which a radius of 4 pixels at the beam centre were excluded. To reduce the influence of the probe profile in the reconstruction, a larger central square region was included in the mask. The final mask used for reconstruction is shown in Fig.~\ref{fgr:SNR diff}(d).

\subsection{Temporal window}
For a time series dataset we collect $n=1...N$ diffraction patterns $I(q,t)$, where in this experiment $N=21,000$. To generate a reconstructable dataset, $I^{R}(q,t)$, from this time-series, we sum together frames using a sliding temporal window of width $x$. Within $I^{R}(q,t)$, the intensity of the $n^{\textnormal{th}}$ diffraction pattern is then given by 

\begin{equation}
    I^{R}_{n}(q,t) = \sum_{i=n-x/2}^{n+x/2} I_{i}(q,t).
\end{equation}

When performing the reconstructions, we excluded frames where $n<x$ and $N-n <x$, ensuring that each reconstructed frame was treated equally.
For the results shown in Fig.~2 in the main text, the final $I^{R}(q,t)$ datasets were generated using temporal window widths of $x=50$ and $x=16$, for the temperatures of 300~K and 340~K, respectively.

\section{XPCS data analysis}

\begin{figure}
    \centering
    \includegraphics[width=\linewidth]{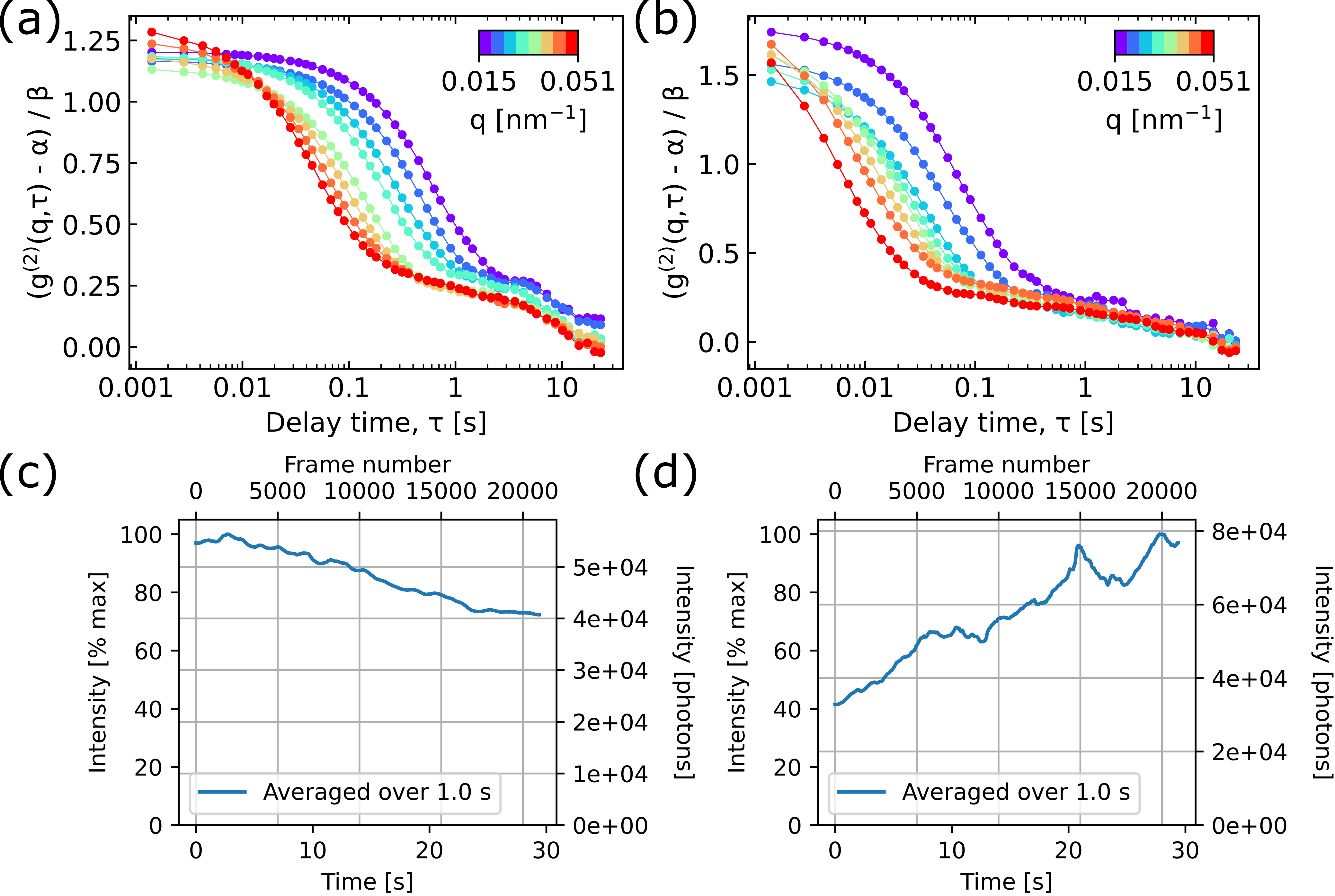}
    \caption{Calculated $g^{(2)}(q,\tau)$ functions as a function of delay time at temperatures of 300~K (a) and 340~K (b) over the whole time series. 
    The colour represents the eight different $q$-partitions used for the analysis, where the partitions are spaced with equal d$q/q$ steps. 
    The $\alpha$ and $\beta$ used for normalisation were obtained from the fits of the data shown in Fig.~1 in the main text.
    The total scattered intensity smoothed over 1~s as a function of time for 300~K (c) and 340~K (d). The change in intensity directly relates to the number of particles within the field-of-view.
    }
    \label{fgr:XPCS full time}
\end{figure}

\begin{figure}
    \centering
    \includegraphics[width=\linewidth]{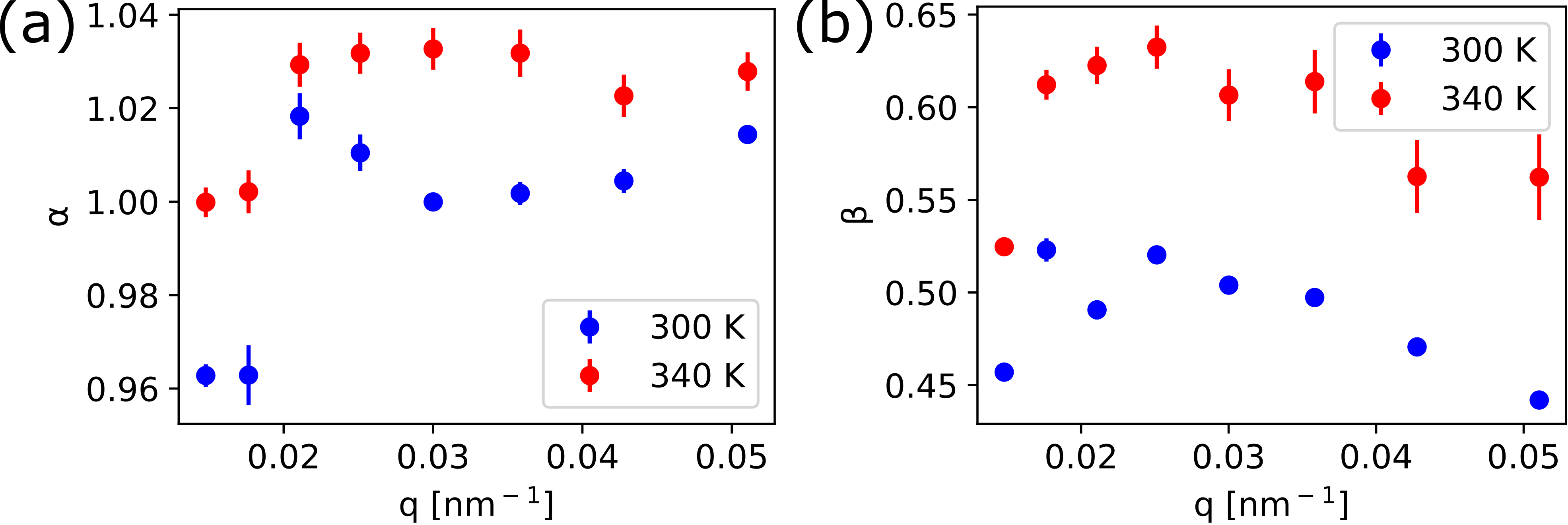}
    \caption{
    The values of (a) the baseline, $\alpha$, and (b) the speckle contrast, $\beta$, obtained by fitting the $g^{(2)}(q,\tau)$ functions shown in Fig.~1 in the main text.
    }
    \label{fgr:XPCS fit param}
\end{figure}

The second-order autocorrelation $g^{(2)}(q,\tau)$ function is calculated by

\begin{equation} \label{eqn:XPCS g2}
    g^{(2)}(q,\tau) = \frac{\left \langle I(q,t) I(q,t+\tau) \right \rangle}
                         {\left \langle I(q,t)  \right \rangle ^2},
\end{equation}

\noindent where $I(q,t)$ is the measured intensity, $q$ is the modulus of the wave-vector, $t$ is time, and $\tau$ is the delay time. 
The results from calculating $g^{(2)}(q,\tau)$ over the whole time-series is shown in Fig.~\ref{fgr:XPCS full time}(a,b), for temperatures of 300~K and 340~K, respectively. 
We see that the $g^{(2)}(q,\tau)$ correlation functions shown in Fig.~\ref{fgr:XPCS full time} exhibit two decays representing fast and slow dynamics.
The fast dynamics are attributed to Brownian motion, which can be fitted using

\begin{equation} \label{eqn:XPCS g2 fit}
    g^{(2)}(q,\tau) = \alpha + \beta  \exp{[-2(\Gamma \tau)^{\gamma}] }  ,
\end{equation}

\noindent where $\alpha$ is the baseline, $\beta$ is the speckle contrast, $\Gamma$ is the relaxation rate, and the exponent $\gamma$ is a measure of the distribution of relaxation times~\cite{lehmkuhler_femtoseconds_2021}.
The final values of $\alpha$ and $\beta$ obtained from the fitting can be seen in Fig.~\ref{fgr:XPCS fit param}, while the values of $\Gamma$ are found in Fig.~1 in the main text, and $\gamma=1$ was used for the fits of both temperatures as we expect the motion to be Brownian.

\begin{figure}
    \centering
    \includegraphics[width=\linewidth]{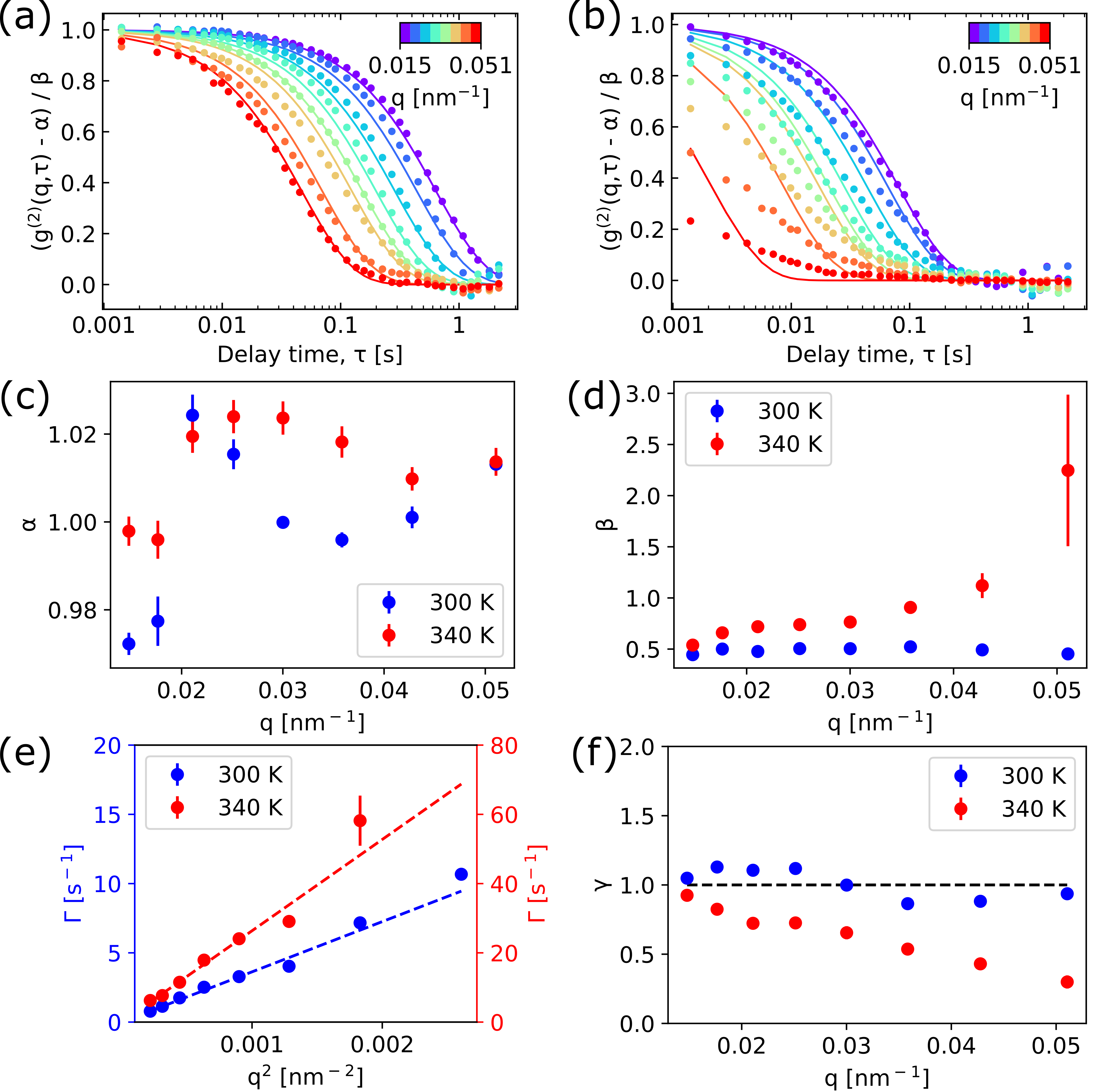}
    \caption{
    Fitting results of both datasets when $\gamma$ is set to be a free parameter. (a,b) Calculated $g^{(2)}(q,\tau)$ functions (points) and their fits (solid lines) as a function of delay time at temperatures of 300~K (a) and 340~K (b) over $3$~s of data near the beginning of the series. 
    The colour represents the eight different $q$-partitions used for the analysis, where the partitions are spaced with equal d$q/q$ steps. 
    The values of (c) the baseline, $\alpha$, (d) the speckle contrast, $\beta$, (e) the relaxation rate, $\Gamma$, and (f) $\gamma$ obtained from the fitting.
    }
    \label{fgr:XPCS fit param gamma 1}
\end{figure}

Results of fits when letting $\gamma$ be a free parameter are shown in Fig.~\ref{fgr:XPCS fit param gamma 1}. We see that the fits at 300~K are all reasonable, with $\gamma \approx 1$ for all $q$-partitions. On the other hand, the fits of the data at 340~K are poor where $\gamma$ constantly decreases at larger $q$, which is not expected for a system undergoing Brownian diffusion. Due to this, we used the result of $\gamma = 1$ at 300~K for the fitting analysis at both temperatures to ensure the analysis of both datasets was identical.

The decay representing slow dynamics we attribute to changes in the scattered intensity, which can be seen in Fig.~\ref{fgr:XPCS full time}(c,d) for 300~K and 340~K data, respectively. Due to this, we limited the fitting analysis to a smaller time range in which the intensity changes remained relatively constant.

The fluctuating intensity is a source of error in the XPCS results and stems from the limitations of the experiment when trying to simultaneously combine CXDI and XPCS. 
In order to simultaneously employ CXDI, the beamsize was reduced to oversample the speckle, and the number of particles was reduced to allow tracking of individual particles, compared to typical XPCS experiments.
A consequence of this is that a change of a single nanoparticle in the field-of-view (FOV) can lead to a significant change in the scattered intensity. 
As XPCS typically employs a larger beamsize and has a larger concentration of particles, the $g^{(2)}(q,\tau)$ correlation curves are insensitive to a few particles entering or leaving the FOV. 
This experiment demonstrates that although CXDI has a more stringent experimental configuration than XPCS which reduces the signal-to-noise ratio (SNR), there is still overlap between an optimal imaging regime for CXDI, and suitable parameters for XPCS.

The $g^{(2)}(q,\tau)$ analysis above calculates the mean correlation between all diffraction patterns of the time series for a given $\tau$. To calculate the correlation between multiple time points, we can use the two-time correlation function, $C(q,t_1,t_2)$, which is calculated as
\begin{equation}
    C(q,t_1,t_2) = \dfrac{\langle I(q,t_1) I(q,t_2)\rangle }
                         {\langle I(q,t_1)\rangle \langle I(q,t_2)\rangle}.
\end{equation}

\begin{figure}
    \centering
    \includegraphics[width=\linewidth]{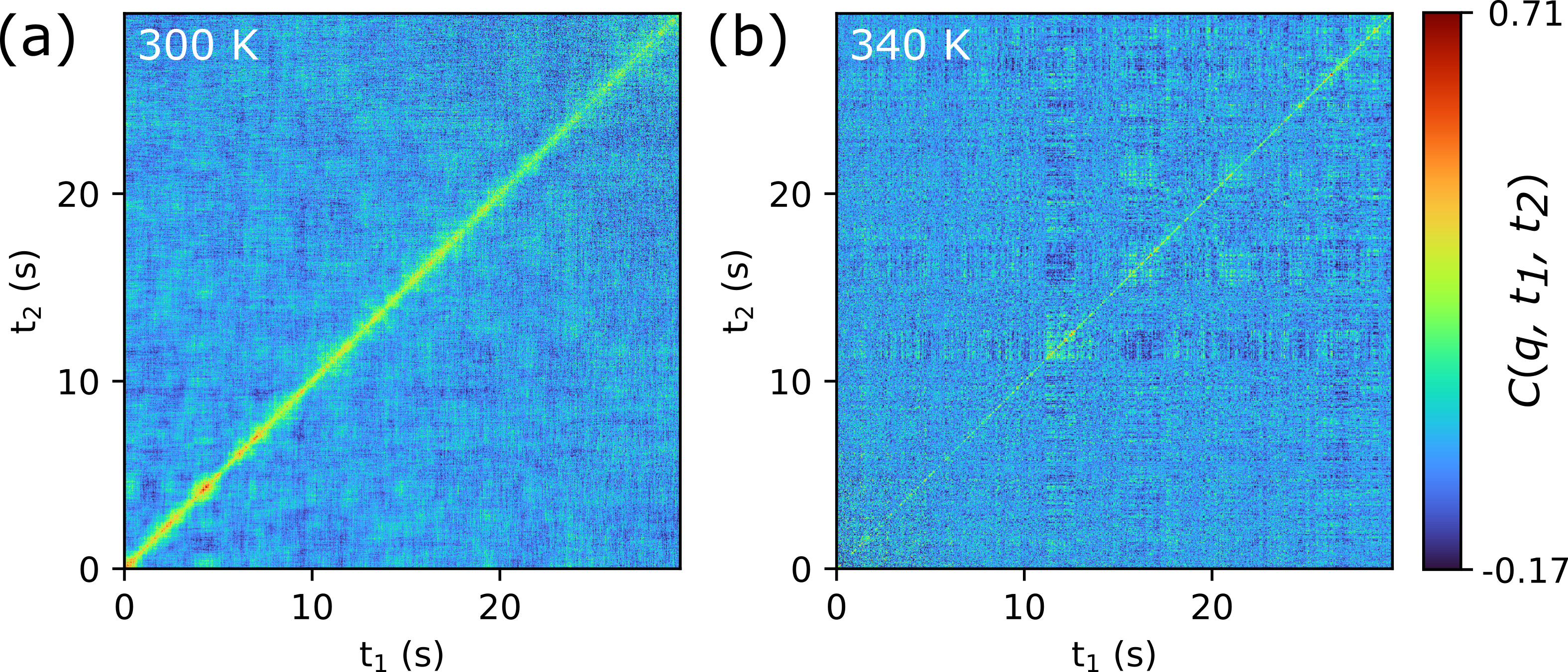}
    \caption{Two time correlation plots at temperatures of 300~K (a) and 340~K (b). The two time correlations were calculated at $q=0.018~$nm$^{-1}$ (300~K) and $q=0.025~$nm$^{-1}$ (340~K).
    }
    \label{fgr:XPCS two time}
\end{figure}

The results from the two-time correlation analysis are shown in Fig.~\ref{fgr:XPCS two time}(a) and (b), for temperatures of 300~K and 340~K, respectively. Although we see the process of agglomeration in the CXDI reconstructions, the two-time correlations appear to be relatively constant with only minor changes to the correlations over time. 

\section{CXDI reconstruction process}

The PyNX package~\cite{favre-nicolin_pynx_2020} was used to perform the iterative algorithm process which consisted of a sequence of 2000 relaxed averaged alternating reflections (RAAR)~\cite{luke_relaxed_2005} + 500 Error Reduction (ER) iterations~\cite{gerchberg_practical_1972}. 
Modulus and support constraints were applied every iteration, with the shrinkwrap algorithm~\cite{marchesini_x-ray_2003} applied every 150 iterations. Pixels within the mask were allowed to fluctuate and were not constrained. Partial coherence correction through the Lucy-Richardson deconvolution~\cite{clark_high-resolution_2012}, as implemented in PyNX, was used. 

Reconstructions were performed serially. The first time point was reconstructed using the autocorrelation as the starting support. For subsequent time points, the final support from the previous time point was used as the starting support.

To begin, data at both temperatures were generated by using a temporal window with length $x=50$ diffraction patterns, and were subsequently reconstructed as described above. 
To reconstruct the data at 340~K at a SNR below the `limit' indicated in the main text, providing reconstructed images with less blurred motion, a new dataset was generated with a temporal window length of $x=16$ diffraction patterns. 
The final supports from the reconstruction of the 50 diffraction pattern data was then used as a starting support for the 16 diffraction pattern data and the supports were then held constant throughout the iterative process.

After all reconstructions were performed, any twin images within the dataset were manually inverted such that the entire dataset reconstructed the same upright object. Images were then aligned using the StackReg~\cite{thevenaz_pyramid_1998} plugin within ImageJ.

\section{Signal-to-Noise Analysis}

Datasets with different SNR values were generated by using temporal windows of different sizes, the smallest being $x=1$ and the largest being $x=750$. 
For a given value of temporal window length $x$, the SNR is calculated as:

\begin{equation} \label{eqn:SNR}
    \textnormal{SNR} =  \left \langle \frac{I^{R}(q,t)}
               {\sqrt{I^{R}(q,t)}}  \right \rangle_n,
\end{equation}

\noindent where $\langle \cdots \rangle _n$ denotes averaging over all diffraction patterns. The uncertainty of the SNR was estimated by the standard deviation.
The SNR was calculated while only ignoring bad pixels and those corresponding to the beamstop within the diffraction pattern, \textit{i.e.}, we did not exclude the central pixels related to the probe in the calculation.

As poor SNR prevented the successful reconstruction of some frames, reconstructions for a given dataset were attempted using PyNX 5 times. By default PyNX uses five tries to reconstruct a dataset per attempt, leading to a possible 25 trials for a given dataset. 
The criteria for success was taken to be convergence of the CDI reconstruction per attempt, not for each trial.

Computational time of the reconstructions was shortened by only reconstructing every 1,000th frame instead of all 21,000 frames, producing 20 unique datasets for a given SNR value. The final values of success percentage are then the average of all 20 datasets.

\section{Particle tracking}

Particle tracking was performed in Python using the TrackPy~\cite{allan_soft-mattertrackpy_2021} package.
The following TrackPy parameters were used to identify particles within each individual frame: diameter = 11~pixels, the minimum integrated brightness = 1000, 
maximum radius-of-gyration of brightness = 10.0, and the minimum separation between features = 4~pixels.
For linking trajectories, the following parameters were used: max displacement = 5~pixels~per~frame, memory = 21000~frames.
Trajectories which were identified for less than 10~frames were ignored.
Drift within the time-series was eliminated using the built-in function within TrackPy before calculating the mean-squared displacement, $\Delta r^2$.
This was then related to the Diffusion coefficient by

\begin{equation} \label{eq:MSD}
    \langle \Delta r^2 \rangle = 4 D \tau.
\end{equation}

\begin{figure}
    \centering
    \includegraphics[width=\linewidth]{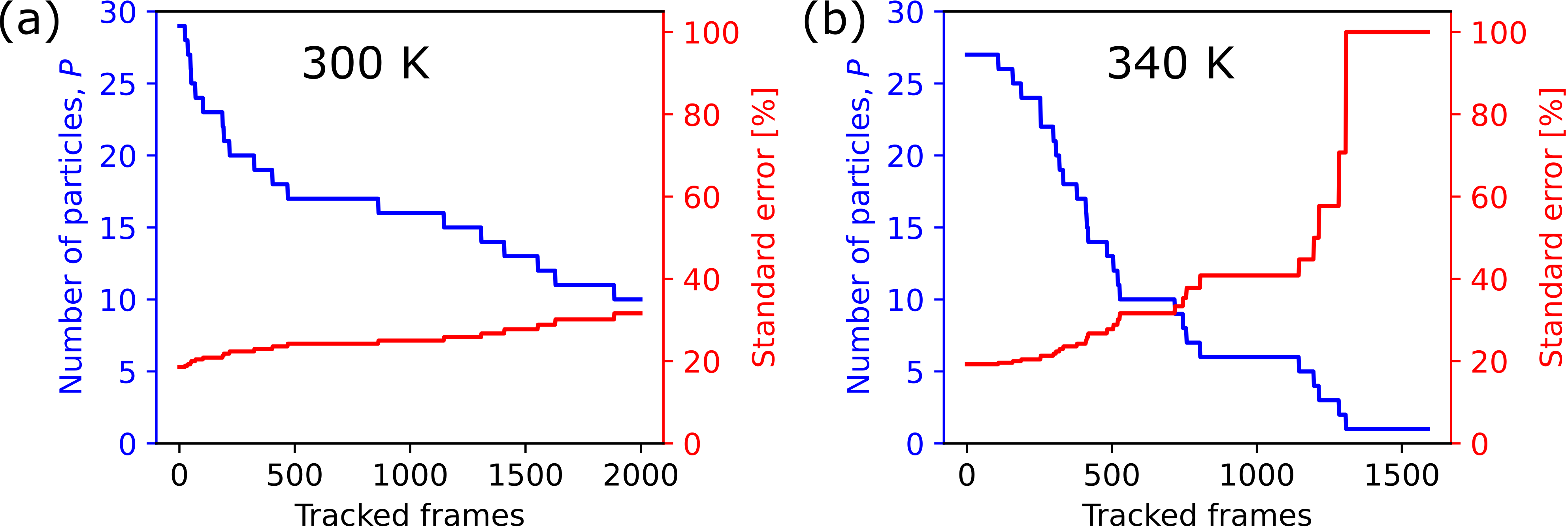}
    \caption{SPT uncertainty using the standard error for data at 300~K (a) and 340~K (b). The blue lines represent the number of particles identified, while the red line is the corresponding standard error. 
    }
    \label{fgr:Tracking error}
\end{figure}

The uncertainty in the SPT results was estimated by the standard error~\cite{catipovic_improving_2013}

\begin{equation}
    \sigma = \dfrac{1}{\sqrt{P}},
\end{equation}

where $\sigma$ is the uncertainty, and $P$ is the number of tracked particles. 
The number of frames for which a particle may be tracked will depend on when a particle enters or leaves the field of view, if there is any overlap with another particle, or also the presence of artefacts in the reconstruction obscuring the ability to identify a particle.
Due to this, the number of tracked particles $P$ decreases for large number of tracked frames, leading to an increase in the uncertainty.
The number of tracked particles, and therefore an estimate of the uncertainty, as a function of tracked frames is shown in Fig~\ref{fgr:Tracking error}(a,b) for the data at 300~K and 340~K, respectively.

To provide a single estimate of the uncertainty on the calculated diffusion coefficients using SPT, we use the mean value of $\sigma$ across all frames. This leads to uncertainties of $\sigma_{\textnormal{300}} = 25.6$\% ($\pm 1,223$~nm$^2$/s), and $\sigma_{\textnormal{340}} = 45.3$\% ($\pm 15,053$~nm$^2$/s). 
We note that the uncertainty in the results at 340~K are quite high.
As the number of particles tracked for more than $1000$ frames for the 340~K case is quite small, it may be more accurate to estimate the uncertainty over only 1000 tracked frames. Doing this results in $\sigma_{\textnormal{340}} = 28.9$\% ($\pm 9,603$~nm$^2$/s), where the agreement between $D^{\textnormal{CXDI}}_{\textnormal{340}}$ and $D^{\textnormal{XPCS}}_{\textnormal{340}}$ is still within error.

\section{Critical sampling frequency}

The critical sampling frequency, $f_c$, refers to the degree of sampling required in order to fully capture the dynamic behaviour of interest. The value of $f_c$ is sample-dependent, and is the inverse of the critical acquisition time, $f_c = 1 / \tau_c$. As described in Ref.~\cite{hinsley_synchrotron_2022}, to estimate a value of $f_c$ for Brownian motion, we can define that the critical acquisition time $\tau_c$ is equal to the time required for the mean displacement of a particle to be half it's diameter, $\Delta r = 0.5 d$. Substituting this into Eq.~\eqref{eq:MSD} we get:

\begin{equation*}
    \Delta r^2 = 4D\tau
    \Rightarrow \left ( \dfrac{d}{2} \right ) ^{2} = \dfrac{4 D}{f_c} ,
\end{equation*}

\noindent and rearranging we get

\begin{equation}
    f_c = \dfrac{16 D}{d^2}.
\end{equation}

For the 300~K data, $D=3,618$~nm$^2$/s, and $f_c=1.45$~Hz, or equivalently a critical acquisition time of $\tau_c = 691$~ms. 
For the 340~K data, $D=28,024$~nm$^2$/s, and $f_c=11.21$~Hz, or equivalently a critical acquisition time of $\tau_c = 89$~ms.

\section{Microrheology analysis}

The viscosity of a liquid can be calculated by

\begin{equation} \label{eqn:viscosity}
    \eta = \frac{k_B T}
             {3 \pi D d}.
\end{equation}

Using $D=3,618$~nm$^2$/s, the XPCS result at 300~K, the viscosity of the solution is $\eta=0.59$~Pa~s.
The composition of the solution is known to be a mixture of water and glycerol, each of which have their own viscosities, $\eta_{W}$ and $\eta_{G}$, respectively.
Following the process in Ref.~\cite{cheng_formula_2008}, both viscosities may be approximated by:

\begin{equation} \label{eqn:viscosity - water}
    \eta_{W} = 12100 \exp{\left[ \frac{T(T-1233)}
                                {9900+70T} \right]},
\end{equation}

\noindent and

\begin{equation} \label{eqn:viscosity - glycerol}
    \eta_{G} = 1.79 \exp{\left[ \frac{T(T-1233)}
                                {36100+360T} \right]}.
\end{equation}

The relative percentage of water can be found through:

\begin{equation} \label{eqn:Water concentration}
    C_W = \frac{\log(\eta / \eta_G)}
               {\log(\eta_W / \eta_G)},
\end{equation}

\noindent and the relative percentage of glycerol can be found through $C_G = 100\% - C_W$. Using the above equations, we obtain $C_W = 12$\% and $C_G = 88$\% at 300~K.

\bibliography{main}

\end{document}